\documentclass[manuscript,screen,nonacm]{acmart}
\usepackage{pifont}
\AtBeginDocument{%
  \providecommand\BibTeX{{%
    \normalfont B\kern-0.5em{\scshape i\kern-0.25em b}\kern-0.8em\TeX}}}

\setcopyright{acmcopyright}
\copyrightyear{2024}
\acmYear{2024}
\acmDOI{XXXXXXX.XXXXXXX}

\acmConference[IUI '24]{29th Annual ACM Conference on Intelligent User Interfaces}{March 18--21,
  2024}{Greenville, SC}
\acmBooktitle{IUI '24: 29th Annual ACM Conference on Intelligent User Interfaces,
 March 18--21, 2024, Greenville, SC} 
\acmPrice{15.00}
\acmISBN{978-1-4503-XXXX-X/18/06}

\begin{document}

\title{QualiGPT: GPT as an easy-to-use tool for qualitative coding}

\author{He Zhang}
\email{hpz5211@psu.edu}
\orcid{0000-0002-8169-1653}
\affiliation{%
  \institution{College of Information Sciences and Technology, Penn State University}
  \city{University Park}
  \state{Pennsylvania}
  \country{USA}
  \postcode{16802}
}
\author{Chuhao Wu}
\email{cjw6297@psu.edu}
\affiliation{%
  \institution{College of Information Sciences and Technology, Penn State University}
  \city{University Park}
  \state{Pennsylvania}
  \country{USA}
  \postcode{16802}
}
\author{Jingyi Xie}
\email{jzx5099@psu.edu}
\affiliation{%
  \institution{College of Information Sciences and Technology, Penn State University}
  \city{University Park}
  \state{Pennsylvania}
  \country{USA}
  \postcode{16802}
}
\author{ChanMin Kim}
\email{cmk604@psu.edu}
\affiliation{%
  \institution{College of Education, Penn State University}
  \city{University Park}
  \state{Pennsylvania}
  \country{USA}
  \postcode{16802}
}
\author{John M. Carroll}
\authornote{Corresponding author.}
\orcid{0000-0001-5189-337X}
\email{jmc56@psu.edu}
\affiliation{%
  \institution{College of Information Sciences and Technology, Penn State University}
  \city{University Park}
  \state{Pennsylvania}
  \country{USA}
  \postcode{16802}
}

\renewcommand{\shortauthors}{Zhang and Wu, et al.}

\begin{abstract}
 Qualitative research delves deeply into individual complex perspectives on technology and various phenomena. However, a meticulous analysis of qualitative data often requires a significant amount of time, especially during the crucial coding stage. Although there is software specifically designed for qualitative evaluation, many of these platforms fall short in terms of automatic coding, intuitive usability, and cost-effectiveness. With the rise of Large Language Models (LLMs) such as GPT-3 and its successors, we are at the forefront of a transformative era for enhancing qualitative analysis. In this paper, we introduce QualiGPT, a specialized tool designed after considering challenges associated with ChatGPT and qualitative analysis. It harnesses the capabilities of the Generative Pretrained Transformer (GPT) and its API for thematic analysis of qualitative data. By comparing traditional manual coding with QualiGPT's analysis on both simulated and actual datasets, we verify that QualiGPT not only refines the qualitative analysis process but also elevates its transparency, credibility, and accessibility. Notably, compared to existing analytical platforms, QualiGPT stands out with its intuitive design, significantly reducing the learning curve and operational barriers for users.
\end{abstract}


\begin{teaserfigure}
  \includegraphics[width=1\linewidth]{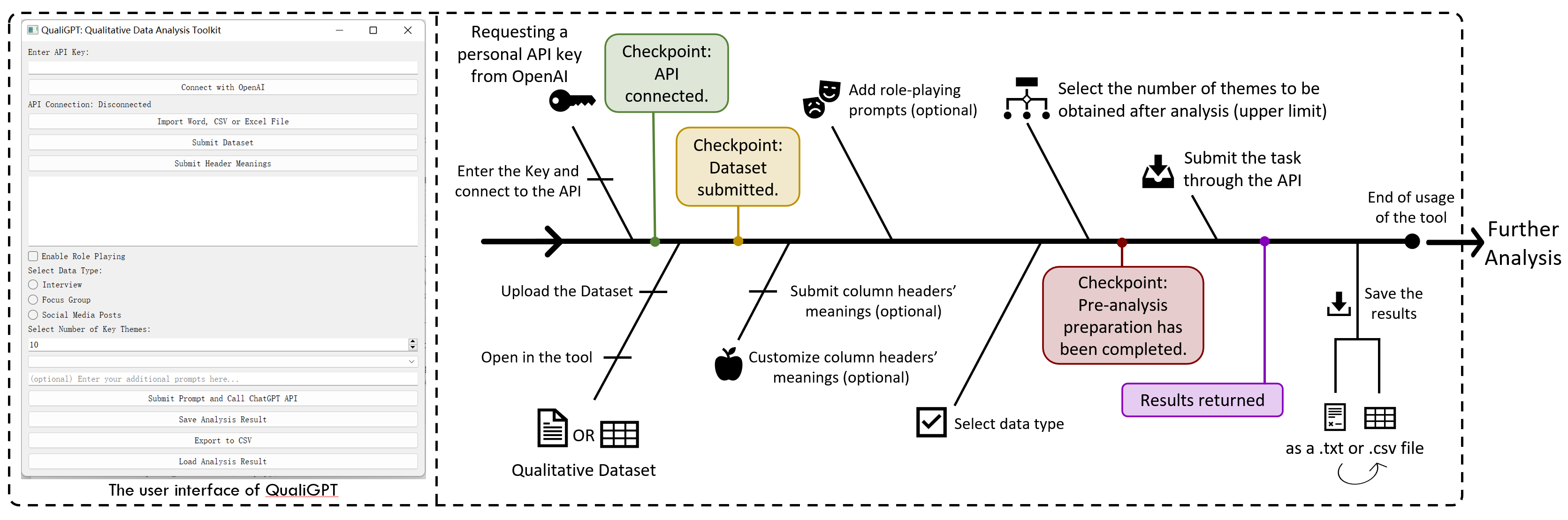}
  \caption{\textbf{Overview of the qualitative analysis toolkit, QualiGPT. The user interface of QualiGPT is displayed on the left. On the right side, the usage flow and design logic of QualiGPT are presented.}}
  \label{fig.teaserfigure}
\end{teaserfigure}


\received{9 October 2023}

\maketitle

\section{Introduction}





Qualitative research provides a unique perspective into individuals' comprehension, attitudes, and insights regarding technology, phenomena, and specific topics. Over time, an increasing number of researchers have acknowledged the significance of qualitative methodologies across diverse fields. However, while these methods are indispensable, analyzing qualitative data can be labor-intensive~\cite{10.1145/3449168}, especially with extensive and complex datasets. Moreover, the task of coding qualitative data not only demands significant effort but also poses challenges related to understanding context and ensuring consistency. Coding, arguably the most crucial task in qualitative analysis, is both a beloved and challenging aspect for analysts. Continuously optimizing methods for processing qualitative data remains a common goal among these professionals. As the production of qualitative data continues to surge, there is an escalating demand for innovative techniques to streamline and enhance the thematic analysis process~\cite{bazeley2013qualitative}.

To address these challenges, researchers have ventured into the development and utilization of qualitative analysis software. These tools employ computer-assisted collaborative efforts to simplify data management and enhance efficiency~\cite{elliott2021exploring}. While such software has indeed streamlined the coding process and improved the quality of coding to some extent, existing platforms like Nvivo\footnote{https://lumivero.com/products/nvivo/} and atlas.ti\footnote{https://atlasti.com/} still have limitations in terms of performance and operational complexity, failing to fully meet the needs of researchers~\cite{hansson2010multiple,bergin2011nvivo}.

In addition to the high subscription costs, learning to use these software tools for qualitative data analysis is not straightforward. Early-career researchers or analysts often find themselves investing a significant amount of time in understanding how to accomplish their target tasks within these environments and navigating the multifaceted UI interfaces~\cite{10.1145/3492844}. However, many of these features are designed to cater to specific needs. In other words, not all functionalities within the software are utilized frequently by analysts, leading to increased learning overheads. As described by Ragavan et al.~\cite{10.1145/3490099.3511161}, analysts not only have to be concerned about their primary tasks at hand but also bear the additional learning costs associated with the tools (software) they choose. Therefore, the development of a more user-friendly tool to reduce the workload of analysts in their primary workflows becomes especially crucial.
Starting from 2022, with the emergence of GPT-3, researchers began to widely recognize the immense potential of Large Language Models (LLMs) in various domains. The subsequent releases of GPT-3.5 and GPT-4, were perceived by many as heralding a comprehensive technological revolution. The advent of large-scale language models seemed to offer a new avenue of possibilities. It was during this time that we were inspired to ponder whether it might be feasible to leverage LLMs to assist in qualitative analysis, aiming to enhance both efficiency and performance. To achieve this objective, we approached it from a practical standpoint, selecting one of the most popular LLM applications developed by OpenAI, ChatGPT and its API, as our research platform to bolster the universality of our research contributions. We drew inspiration from the recent work of Zhang et al.~\cite{zhang2023redefining} on enhancing qualitative analysis using ChatGPT, emphasizing the importance of prompts and by extending some of the future work they had highlighted.

In summary, this study first categorizes and summarizes the typical issues encountered when using ChatGPT, identifying four major categories that encompass eight common types of erroneous ChatGPT responses. Concurrently, we compiled concerns from previous studies wherein analysts expressed reservations about employing ChatGPT for qualitative analysis tasks, as well as the challenges ChatGPT faces in such contexts. With these issues and challenges in mind, we introduced QualiGPT: a user-friendly integrated tool built on API and prompt design, specifically tailored for thematic analysis of qualitative data.

We deployed QualiGPT on both simulated and real datasets and compared its performance to manual coding. The results show that this tool effectively addresses the challenges inherent in the traditional qualitative data coding process. It streamlines the qualitative analysis workflow, reduces costs associated with processing qualitative data, and alleviates concerns regarding transparency and credibility in using ChatGPT for qualitative analysis. Additionally, due to its integrated design and API implementation, QualiGPT offers marked improvements in usability, user-friendliness, privacy protection, and performance over the web version of ChatGPT. When compared to conventional software, QualiGPT provides a more insightful user interface, significantly lowering the learning and usage costs for researchers.

%
%

\section{Background and Related Work}

\subsection{Thematic Analysis of Qualitative Data}

Thematic analysis is a method used to encode qualitative data gathered from various sources, such as interviews, focus groups, social platforms, or field research. Its primary objective is to identify patterns or themes that aptly describe and organize the observed data or interpret various facets of a phenomenon~\cite{braun2012thematic,boyatzis1998transforming}. This technique has been applied in diverse fields, including psychology~\cite{braun2006using,willig2013ebook}, sociology~\cite{guest2011applied}, education~\cite{xu2020applying,oliveira2021exploratory}, and health and well-being~\cite{braun2014can,crawford2008professional,chapman2015qualitative}.

Two primary approaches to thematic analysis have emerged in existing literature, each offering distinct perspectives on theme identification~\cite{braun2006using,braun2012thematic}. The inductive approach emphasizes deriving meaning directly from the data, free from the constraints of prior knowledge or pre-existing theories~\cite{fereday2006demonstrating,varpio2020distinctions}. In contrast, the deductive approach leverages existing theories or frameworks to pinpoint specific themes of interest~\cite{braun2012thematic,fereday2006demonstrating,pearse2019illustration}.

In our research, we concentrated on the inductive, data-driven thematic analysis approach. We have developed a toolkit allowing researchers to automatically analyze qualitative data using LLMs.

Braun and Clarke's 6-phase coding framework~\cite{braun2006using,clarke2013teaching} is widely adopted in thematic analysis, regardless of the specific approach taken~\cite{braun2012thematic}. This framework delineates a process that starts with data familiarization. It then progresses to code generation, combines codes into themes, reviews those themes, determines their significance, and finally culminates in the preparation of the final report. Inherently iterative, this framework often compels researchers to revisit earlier steps when faced with new data or emerging themes that necessitate further exploration. Such an approach can yield more nuanced and thorough thematic outcomes.


Thematic analysis serves as a potent tool for qualitative data, yet its application is not devoid of challenges inherent to the intricacies of qualitative research~\cite{10.1145/3449168}.

\textit{First}, the challenge of ``researcher subjectivity'' arises. Each researcher brings their unique biases to data interpretation, which can lead to the emergence of diverse themes from the same dataset~\cite{799955}. This variability introduces replicability concerns~\cite{braun2019reflecting,10.1145/3544548.3580766}. Consequently, it's crucial to uphold transparency and credibility of the results~\cite{castleberry2018thematic} and to detail the interpretive process comprehensively.

\textit{Second}, the resource-intensive nature of thematic analysis stands as a considerable challenge, especially when dealing with extensive datasets~\cite{castleberry2018thematic,guest2013collecting}. Recognizing patterns and themes demands deep engagement, necessitating significant time and effort, extending beyond mere data coding~\cite{terry2017thematic}.

\textit{Third}, questions of generalizability come to the fore~\cite{leung2015validity}. While thematic analysis offers profound insights into a specific context, extrapolating these insights to varying contexts can be restrictive, thereby limiting the generalizability of outcomes.

\textit{Fourth}, the caliber of the collected data forms the bedrock of thematic analysis' robustness~\cite{braun2021saturate}. The emergence of significant and relevant themes hinges on the depth and accuracy of the data. Shortcomings in data collection can undermine the validity and richness of the themes derived.

This study explores the extent to which AI's integration into thematic analysis can address some of these challenges. Specifically, we examine how AI can provide efficient resource management, tackling the ``resource-intensive'' nature of thematic analysis.

\subsection{Prompt Engineering}

Prompt engineering is the deliberate design and optimization of instructions, or ``prompts'', aimed at enhancing the performance and accuracy of LLMs when generating outputs~\cite{reynolds2021prompt,zamfirescu2023johnny}. This strategy is crucial as the type and specificity of prompts provided to LLMs can significantly shape their responses.

ChatGPT by OpenAI, developed within the Generative Pretrained Transformer (GPT) framework, underscores the importance of prompt engineering~\cite{fiannaca2023Programming}. Renowned for its expertise in diverse language tasks, such as producing human-like text, content generation, sentence completion, and in-depth essay or report writing~\cite{liebrenz2023generating,alkaissi2023artificial,bishop2023computer,macdonald2023can}, ChatGPT is not immune to errors. It may yield outputs that seem nonsensical or incorrect, particularly when faced with unclear or ambiguous prompts~\cite{shen2023chatgpt,hassani2023role}.

The value of prompt engineering gains further emphasis from studies revealing improved outcomes when LLMs like ChatGPT receive meticulously crafted prompts. Techniques such as few-shot learning~\cite{zhao2021Calibrate}, chain-of-thought methods~\cite{wei2022ChainofThought}, and role-playing scenarios~\cite{gao2023Prompt} have demonstrated considerable efficacy. However, the performance of ChatGPT, even when paired with refined prompt engineering, can differ based on the domain in question. Mastery in domain-specific knowledge is pivotal for honing the model's efficacy~\cite{tian2023Opportunities, wang2023Brief}. Thus, practitioners are encouraged to weigh the specific application context carefully during prompt engineering~\cite{heston2023Prompt}.
For areas like qualitative analysis, employing an iterative methodology—consistently adapting and evaluating diverse prompt engineering strategies—may be instrumental in harnessing the full potential of ChatGPT.

\subsection{Computer-Assisted Qualitative Data Analysis}
With the increasing amount of qualitative data being generated, Computer-Assisted Qualitative Data Analysis (CAQDA) has been playing a critical role in qualitative research. Over the past decades, a large number of CAQDA applications and software has emerged to help researchers organize, manage, and analyze data. The history of CAQDA can be traced back to the introduction of computers in the 1980s \cite{pilkington1996Use}, and \citet{weitzman1995computer} have categorized 24 software programs available at that time into 5 types. \citet{chandra2019ComputerAssisted,sanchez-gomez2019Evaluation} provide detailed introduction of contemporary CAQDA software and highlight that the choice of program must consider the nature of data, type of coding, and other factors during the research. CAQDA software nowadays offer a wide range of functionalities such as processing data from multiple media channels (text, picture, audio, and video), visualizing the analysis results through automatic plotting of data, and quickly generating predefined and customized reports \cite{phillips2018quick}. Some applications such as Dedoose focuses on facilitating collaborative QDA by enabling real-time data exchange among multiple researchers \cite{huynh2021Media}. While the capabilities vary greatly among applications, more advanced functionalities often comes at the cost of a high subscription fee \cite{renaissancerachel202315} that potentially deter researchers away. As a result, some free and open source alternatives have been developed to support the growing need of qualitative research, such as Taguette \cite{rampin2021Taguette} and RQDA \cite{chandra2019overview}, although their functionalities tend to be more basic than commercial products. Another problem with CAQDA software is their user experience and learnability. \citet{paulus2013Digital} find that initial encounters can be intimidating for novices, yet with proper guidance researchers can effectively integrate these tools. Still, studies on the interface design of CAQDA are rather limited and a comparison of both commercial and open source applications is necessary for designing better tools. 

\subsection{AI in Qualitative Research}
The combination of artificial intelligence (AI) and qualitative research has begun to redefine how researchers approach qualitative data and analysis~\cite{haque2022think,10.1145/3581754.3584136}. Technologies, especially AI algorithms, provide potential for improved efficiency in analyzing large datasets, a task that traditionally requires substantial time and resources when conducted by human analysts. In fact, in earlier years, researchers have been using computers or technologies to assist in qualitative studies~\cite{coffey1996qualitative,weitzman1995computer,thunberg2022pioneering}. 

AI can be used to gather and organize qualitative data from various sources, like social media platforms, online forums, and digital archives. This not only saves time and resources but can also uncover a wider range of data points that might be overlooked in manual collection~\cite{feng2023investigating}. Also, AI-powered transcription services can transcribe audio and video data into text format quickly and accurately. Typically, transcription and encoding in qualitative research present the biggest challenges for researchers, often consuming a lot of time. However, a good assistant tool allows researchers to focus more on analysis rather than on data preparation~\cite{10.1145/3173574.3173922}. AI models can provide initial analysis of textual data by summarizing content, identifying key themes, sentiments, or trends, and even insightful advice and generating questions that can help guide further research~\cite{cui2023survey,ma2020machine,khurana2023natural,shaik2022review,10.1145/3411764.3445591}. By comparing AI findings with human analysis, researchers can increase the validity and reliability of their findings~\cite{10.1145/3544548.3581352}. With AI's ability to process data rapidly, researchers can conduct real-time analysis during data collection, helping them adjust their research approach as needed based on preliminary findings~\cite{panda2019artificial}. The advent of automated qualitative analysis techniques has enabled qualitative researchers to analyze volumes of data that would be difficult to analyze manually~\cite{welsh2002dealing}, and the rise of LLMs may further enhance the efficiency of analysis.

Despite the impressive capabilities of AI, machine learning (ML), and LLMs, the complex nature of qualitative analysis presents unique challenges that these technologies are still learning to navigate~\cite{10.1145/3544548.3580688}.

\section{Comparison of Existing Qualitative Analysis Software}

In this section, we compare the capabilities of QualiGPT with widely used qualitative analysis software tools, including NVivo\footnote{https://lumivero.com/products/nvivo/}, 
Atlas.ti\footnote{https://atlasti.com/}, 
MAXQDA\footnote{https://www.maxqda.com/}, and
Dedoose\footnote{https://www.dedoose.com/}. 
These tools aid in the organization, coding, and analysis of qualitative data. 

\subsection{User Interface - Cost of Learning}
\begin{figure}[ht]
  \centering
  \includegraphics[width=1\linewidth]{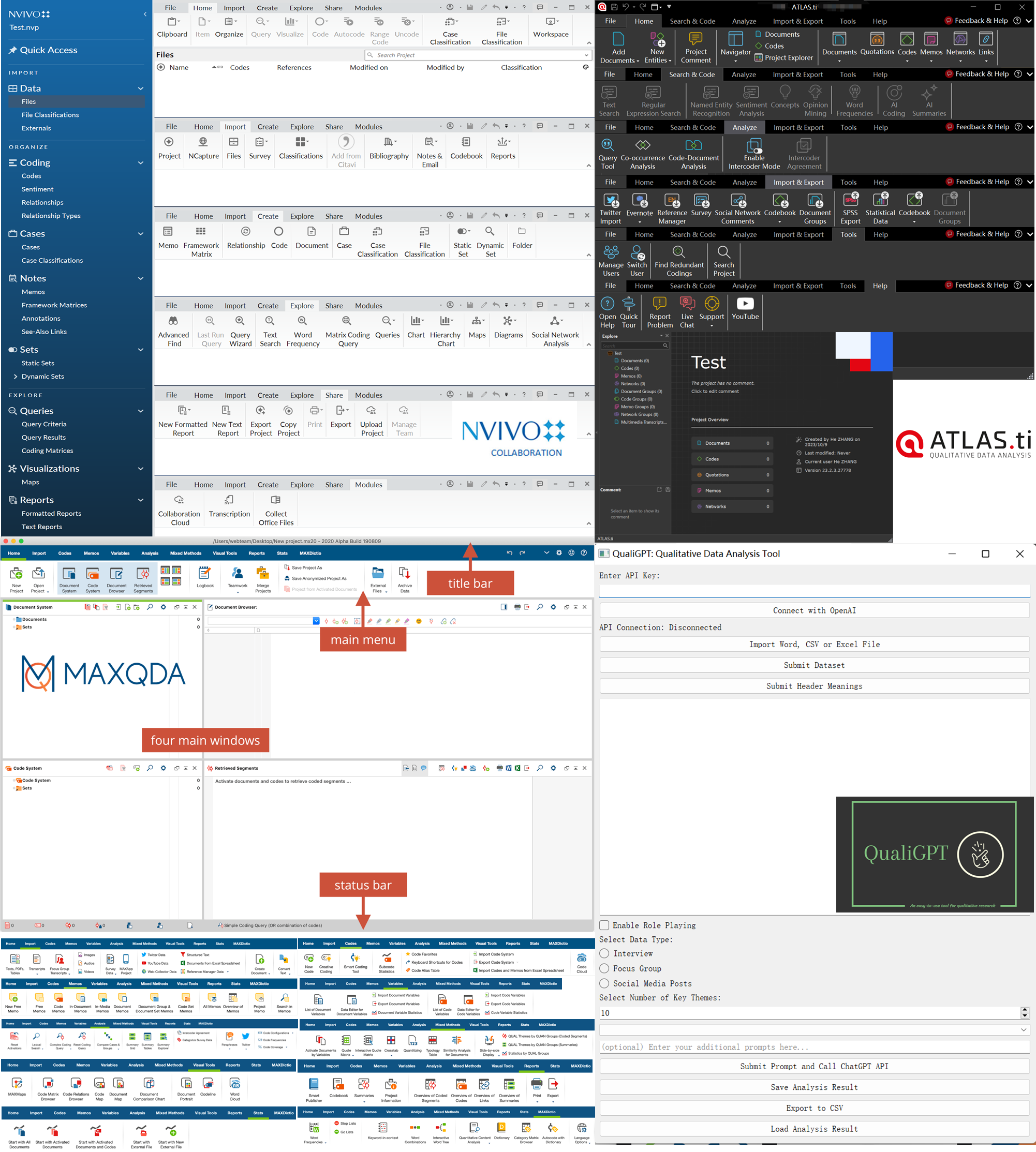}
  \caption{\textbf{Software User Interface Comparison. The top-left showcases Nvivo 14, the top-right displays Atlas.ti, the bottom-left features MAXQDA, and the bottom-right presents QualiGPT.}}
  \label{fig.softwares}
\end{figure}

We compared the user interfaces of three mainstream qualitative analysis software solutions with QualiGPT, as shown in Fig.~\ref{fig.softwares}. It's evident that these commercial software solutions offer a plethora of features, providing users with a wide range of choices. However, this is a double-edged sword. A complex user interface increases the learning curve for users. Typically, users need to undergo extensive training to proficiently use these commercial tools. Moreover, intricate interaction logic can make the analysis process lengthy and prone to errors due to improper operations (using features in the wrong sequence or manner). In contrast, QualiGPT offers a streamlined operational approach, as depicted in Fig.~\ref{fig.teaserfigure}, significantly reducing the potential for errors and time costs due to unfamiliarity with the software.

\subsection{Collaborative Coding}

Both NVivo and Atlas.ti allow researchers to work on separate parts of a project and later merge their work, with cloud services enabling more synchronized collaboration. MAXQDA also facilitates teamwork by allowing independent coding which can be merged and compared for consistency. Being cloud-based platform, Dedoose inherently excels in real-time collaboration, enabling multiple users to simultaneously access, code, and analyze data. 

In contrast, ChatGPT does not inherently support collaborative coding and lacks built-in features for multiple users to collaborate in real-time or asynchronously. However, QualiGPT, as a tool, surpasses mainstream commercial qualitative analysis software in terms of learning curve and coding speed. This allows researchers to quickly code using this tool, saving more time for protocol coding and subsequent analysis. Furthermore, we propose the concept of using QualiGPT as a co-researcher. This implies that qualitative analysts can consider it as an additional independent coder, engaging in discussions with QualiGPT to gain deeper insights.

\subsection{Natural Language Processing Capabilities}

Traditional tools such as NVivo, Atlas.ti, MAXQDA, and Dedoose, are designed primarily for manual qualitative data coding and analysis. They incorporate some basic Natural Language Processing (NLP) features, such as text search, word frequency analysis, and auto-coding based on keyword recognition.

NVivo, for example, offers features like text sentiment analysis and automated insights which utilize underlying NLP principles. Atlas.ti and MAXQDA focus more on manual coding but provide powerful text search and retrieval features. Dedoose, being cloud-based, emphasizes ease of use and collaboration but has limited advanced NLP functionalities compared to dedicated NLP models. 

In contrast, ChatGPT stands out primarily as an advanced application based on LLMs. It's designed to understand and generate text at a near-human level. It can process and respond to prompts dynamically, and its strength lies in generating coherent, contextually relevant text based on extensive training data. 

\subsection{Licensing and Pricing}
Licensing and pricing structures among qualitative software tools and AI models exhibit certain commonalities. Traditional qualitative tools like NVivo, Atlas.ti, and MAXQDA offer both perpetual and subscription-based licenses, with varied pricing tiers to accommodate students, academics, and commercial users. As of October 2023, the pricing for NVivo 14 for commercial user purposes is \$2,038.00, and for students, the subscription cost is \$118 every 12 months. Atlas.ti charges \$666 annually for a single desktop commercial user, while the lowest pricing for student desktop software is \$51 every 6 months. In addition, for other features like advanced collaboration, commercial software typically has additional subscription fees, such as the NVivo Collaboration Cloud subscription which costs \$499.00. Cloud-based platform, Dedoose, predominantly operates on subscription models, adjusting prices based on usage limits, project count, or data storage needs. Similarly, the standard version of ChatGPT is free for registered users, while the premium version (using GPT-4's ChatGPT Plus) is available on a monthly subscription basis at \$20/month\footnote{https://openai.com/blog/chatgpt-plus}. For the API, OpenAI charges based on usage volume (with GPT-3.5 Turbo priced at a minimum of \$0.0015 per 1K tokens, and GPT-4 priced at a minimum of \$0.03 per 1K tokens\footnote{https://openai.com/pricing}). Currently, the 'Moore's Law' of LLMs, known as scaling laws~\cite{kaplan2020scaling}, is beginning to take effect. As model sizes, dataset dimensions, and computational capabilities increase, the performance of the models is expected to improve. Concurrently, the cost of invoking the GPT API may also become more affordable~\cite{Masood_2023}.

\section{Overall Motivation and Design Considerations of QualiGPT}
Our initial motivation stemmed from the real experiences of researchers. As described in the introduction, currently, applying LLMs qualitative analysis may help alleviate the burden on researchers. Based on practical experiences and prior literature, we identified the shortcomings and typical errors of using ChatGPT. This further reinforced our conviction to design an integrated tool. 

By leveraging techniques proposed by researchers for using ChatGPT in qualitative task analysis, we tested and refined these techniques on the web version of ChatGPT. We integrated the solutions into our toolkit, serving as resources and prior knowledge for the development of QualiGPT. Specifically, the design considerations encompass two main parts: the first pertains to the common concerns of qualitative analysts about applying ChatGPT to qualitative analysis tasks, as introduced in Section~\ref{challengesofchatgpt}. The second pertains to some of the current shortcomings of the web version of ChatGPT, as discussed in Section~\ref{initialperformance}.

\subsection{Common Challenges and Concerns of Qualitative Analysis and the Use of ChatGPT in the Qualitative Analysis Process}\label{challengesofchatgpt}
In the study by Zhang et al.~\cite{zhang2023redefining}, they pinpointed several challenges of incorporating LLMs into the qualitative data analysis process through in-depth interviews with qualitative researchers. We revisited these challenges and further contemplated how to address them by designing an integrated tool.

\subsubsection{Lack of Transparency}
One of the primary concerns that deter qualitative researchers from embracing analysis methods or auxiliary tools like ChatGPT is the issue of transparency in data processing~\cite{DWIVEDI2023102642}. This mirrors a common challenge faced by many artificial intelligence technologies, often referred to as the 'black box' problem~\cite{8466590}. Most users of LLM applications remain in the dark about how their requests are processed and fulfilled. Even the developers of LLMs can find it challenging to discern the intricacies within complex neural network training. The transparency issue with ChatGPT is essentially a reflection of concerns about the interpretability of artificial intelligence. However, while researchers express concerns about interpretability, in many cases, the high performance offered by these technologies and applications has brought irreplaceable value to research work or everyday life. Therefore, enhancing transparency and interpretability when collaborating with such technologies becomes paramount~\cite{10.1145/3538392}. Fortunately, as an interactive AI application, ChatGPT allows us, from a user's perspective, to indirectly influence its behavior. While we can't directly control model parameters and architecture, we can guide ChatGPT to self-explain by improving the quality of prompt engineering. Specifically, when applying ChatGPT to qualitative data analysis, we address this issue by designing more explicit prompts, prompting ChatGPT to provide more interpretable responses. For the coding of qualitative data, the codes should be derived from the actual qualitative data. Thus, when researchers can inspect the results, it undoubtedly boosts their confidence. Therefore, requiring ChatGPT to reference the original data for its analysis results is essential, and this can be achieved through prompt design.

\subsubsection{Consistency Issues of ChatGPT}
ChatGPT has been reported to exhibit some consistency issues in its outputs. 
\begin{itemize}
  \item \textbf{Consistency issues:} The consistency of ChatGPT's outputs is highly dependent on the input prompts; different phrasings can lead to varied answers. Therefore, unified, highly usable, and standardized prompts are crucial for ChatGPT to effectively complete tasks.
  \item \textbf{Lack of understanding context:} One of the reasons for the inconsistency in ChatGPT's outputs is its memory issues in multi-turn dialogues. While ChatGPT's conversational capability is acknowledged as an advantage by users, in multi-turn dialogues, ChatGPT may "forget" previous inputs and outputs, leading to inconsistent or contradictory subsequent outputs. Especially when users input multiple, complex prompts consecutively, ChatGPT is more likely to consider different prompts in isolation rather than analyzing them in the context of the conversation. Thus, precise prompts and consolidating multiple requirements into a single prompt can enhance the performance of ChatGPT's responses.
  \item \textbf{Broad or vague responses:} To avoid providing incorrect answers, ChatGPT might produce overly broad or vague responses. This issue can typically be mitigated by refining the quality of prompts. However, evaluating the performance of this issue in qualitative research is challenging, as it often requires comprehensive manual coding, which contradicts the original intent of our tool design. Therefore, we revisited the needs of researchers and ultimately positioned the tool to assist in the coding work of qualitative data tasks. This allows researchers to gain initial insights from the results and draw inspiration for subsequent research, serving as a method to reduce the preliminary workload and benefit researchers. Frankly, this preliminary work (coding) still constitutes one of the most labor-intensive portions of the entire qualitative analysis process.
  \item \textbf{Lack of Fixed Perspective and Absence of Reproducibility:} ChatGPT generates answers based on its training data, but it's challenging to assert its subjectivity, meaning it doesn't think like humans. ChatGPT doesn't have fixed ''beliefs'' or ''views'', which can lead it to express contradictory opinions on certain matters. Simply put, the responses generated by ChatGPT can vary each time. However, ChatGPT can still offer insights by ''reading'' the data. Previous research has shown that using ChatGPT for analysis during the qualitative data coding phase can provide insights similar to those of researchers. These insights can inspire qualitative researchers and enhance their understanding of the qualitative data. To address the issue of reproducibility, setting more precise and highly formatted prompts can effectively reduce the randomness in the format of ChatGPT's output responses. In the qualitative data coding process, regarding the randomness of ChatGPT's generated content, users can mitigate this by having ChatGPT generate a broader range of themes, providing more flexible redundancy in subsequent target theme selections
\end{itemize}

\subsubsection{Designing Prompts is Difficult and Time-consuming}
An increasing number of users are intrigued by the potential capabilities of large language model applications, such as ChatGPT. However, crafting appropriate prompts is not a straightforward task. Currently, the internet is replete with tutorials and methods on how to use prompts. The quality of these resources varies widely, and there hasn't been a standardized rule for prompt design established within the user community. While ChatGPT's conversational interaction mode and openness to input content allow users to interact with it using almost any prompt, this approach is not ideal for formal research. The prompt design framework proposed by Zhang et al.~\cite{zhang2023redefining} for qualitative analysis tasks has greatly facilitated the use of ChatGPT for standardized tasks. Prompts designed based on this framework define the format of input data and expected output results, the methods for data processing and analysis, and considerations for the interpretability and significance of the output results. They also incorporate additional designs (such as role-playing and friendly dialogue) to enhance the quality of ChatGPT's outputs. However, we found that while this prompt design framework offers valuable insights, it still requires users to craft their prompts. Even though users can easily design prompts step-by-step based on the framework, it remains time-consuming. Therefore, we adopted a similar prompt design approach and stored the designed prompts as presets integrated into QualiGPT. This allows users to quickly invoke preset prompts based on their specific needs, significantly reducing the workload in the prompt design phase and enhancing user interaction through an intuitive visual interface.

\subsubsection{Challenges in Understanding ChatGPT's Responses:} The time taken to read, understand, and evaluate ChatGPT's results is not necessarily shorter than the original workflow. This is understandable, as users still need to comprehend the content provided by ChatGPT through reading. If this content is voluminous and disorganized, understanding it might be as challenging as the coding process for the original data, which contradicts our initial intent of leveraging ChatGPT to enhance the efficiency of qualitative analysis. On one hand, formatted content with high readability can improve reading efficiency, so we can design prompts to standardize ChatGPT's outputs. On the other hand, adopting appropriate strategies to prioritize reading sequences can help individuals quickly grasp more common or crucial concepts.

\subsubsection{Data Privacy and Security:} In today's digital age, concerns about information security, particularly data privacy, affect virtually everyone to varying degrees. When using large language model applications to process research data, a primary concern for researchers is the potential leakage of sensitive information. Historically, the impact of data breaches can be catastrophic~\cite{6246173}. While researchers, corporations, and governmental bodies in relevant fields continue to explore and refine methods to protect user data privacy, such as the GDPR~\cite{voigt2017eu} and various encryption techniques~\cite{salavi2019survey}, it's equally vital to consider how ordinary users or non-specialists can enhance their privacy protection online. The efforts we can make individually might be limited, so it becomes crucial to strengthen data privacy protection using publicly available information and methods. To this end, we reviewed OpenAI's policies~\footnote{https://openai.com/security. Accessed October 7, 2023.} related to privacy~\footnote{https://openai.com/enterprise-privacy. Accessed October 7, 2023.}. The findings indicate that, for ChatGPT as a non-API consumer application, data submitted by users can be used to improve the model. However, for the API service, the data is not used for model training. Given this, trusting the service based on its policies and public information, it's evident that using the API service is a better choice from a data security perspective~\cite{mather2009cloud}.

\subsection{ChatGPT Initial Performance Test}\label{initialperformance}
Beyond the issues mentioned before, in this study, we also conducted real-world testing on the web version of ChatGPT. We extracted a small subset of data (1,000 entries from a public Discord channel) from a real-world social media dataset. After removing sensitive information, we tested it using the web versions of ChatGPT-3.5 and ChatGPT-4. The tests aimed to assess whether ChatGPT could return correct results, produce content in a standardized format, and ensure the accuracy of the content. The results revealed that when using detailed and formatted prompts to instruct ChatGPT to perform specific, refined tasks, there was minimal performance difference between ChatGPT-3.5 and ChatGPT-4. This aligns with the findings of Zhang et al.~\cite{zhang2023redefining} Given the current pricing structure of OpenAI~\footnote{https://openai.com/pricing. As of October 7, 2023, the cost of invoking the GPT-4 model for every 1K tokens via the API is more than 10 times higher than that of invoking GPT-3.5 Turbo.}, we believe that using GPT-3.5 offers a better cost-benefit ratio. 

\subsection{Typical Errors: Bad Responds from ChatGPT}
We also summarized typical errors, examples, and potential solutions of responds from ChatGPT encountered during the testing process, as shown in Table~\ref{tab:ErrorMessage}. These primarily include (1) network errors, (2) incorrect handling of data, (3) Violation of policy, and (4) Out of limits. It's worth noting that these errors are not exclusive to using ChatGPT for qualitative analysis tasks and have a certain universality.

\begin{table}[ht]
\centering
\resizebox{\columnwidth}{!}{%
\begin{tabular}{lcccccr}
\hline
Error Message in ChatGPT & Error Type & Error Descriptions & Example of Input & \begin{tabular}[c]{@{}c@{}}Example \\ of \\ Error Output\end{tabular} & R* & Solution \\ \hline
"Network errors" & \begin{tabular}[c]{@{}c@{}}Network \\ connection\end{tabular} & Network errors & Any input & \begin{tabular}[c]{@{}c@{}}Return \\ error message\end{tabular} & Y & Refresh and regenerate \\
\begin{tabular}[c]{@{}l@{}}"Something went wrong. \\ If this issue persists please \\ contact us through our help \\ center at help.openai.com"\end{tabular} & \begin{tabular}[c]{@{}c@{}}Network \\ connection\end{tabular} & Not processed errors & Any input & \begin{tabular}[c]{@{}c@{}}Return \\ error message\end{tabular} & Y & Refresh and regenerate \\
\begin{tabular}[c]{@{}l@{}}"This content may violate \\ our content policy. If you \\ believe this to be in error, \\ please submit your feedback \\ — your input will aid our \\ research in this area."\end{tabular} & \begin{tabular}[c]{@{}c@{}}Violation \\ of policy, \\ because of\\ the forbidden \\ word\end{tabular} & \begin{tabular}[c]{@{}c@{}}This is because ChatGPT \\ detects/mistakenly believes \\ that the input contains \\ information that violates \\ the content policy.\end{tabular} & \begin{tabular}[c]{@{}c@{}}"Because I'm a \\ **13 YEAR OLD KID** \\ and I was live on twitch \\ yesterday then all of \\ a sudden some perverted, \\ creepy pedophile just joins \\ my stream and starts \\ to sexually harass me"\end{tabular} & \begin{tabular}[c]{@{}c@{}}Return \\ error message\end{tabular} & Y & \begin{tabular}[c]{@{}r@{}}Clarify the prompt \\ so that ChatGPT \\ understands that this is \\ not a request, but \\ something that needs \\ to be analyzed.\end{tabular} \\
\begin{tabular}[c]{@{}l@{}}"The message you submitted \\ was too long, please reload \\ the conversation and submit \\ something shorter."\end{tabular} & Out of limits & \begin{tabular}[c]{@{}c@{}}Out of the maximum number\\  of tokens that can be entered \\ in a single entry.\end{tabular} & Over 4096 tokens per entry & \begin{tabular}[c]{@{}c@{}}Return \\ error message\end{tabular} & Y & \begin{tabular}[c]{@{}r@{}}Input the content \\ into ChatGPT \\ in separate sections.\end{tabular} \\
\begin{tabular}[c]{@{}l@{}}"Only one message at a time. \\ Please allow any other \\ responses to complete \\ before sending another \\ message, or wait one minute."\end{tabular} & Out of limits & \begin{tabular}[c]{@{}c@{}}System crashes due to \\ network latency \\ or frequent inputs.\\ ChatGPT does not allow \\ multi-threading \\ per chat window.\end{tabular} & \begin{tabular}[c]{@{}c@{}}New content was inputted \\ when ChatGPT had not \\ provided a complete output \\ (either because it was \\ still in progress or had \\ stopped generating early)."\end{tabular} & \begin{tabular}[c]{@{}c@{}}Return \\ error message\end{tabular} & Y & \begin{tabular}[c]{@{}r@{}}Wait for a while \\ and regenerate\end{tabular} \\
\begin{tabular}[c]{@{}l@{}}"I'm sorry, but I won't be able \\ to assist with that request" \\ or \\ "I'm sorry, but I can't assist \\ with that request."\end{tabular} & \begin{tabular}[c]{@{}c@{}}Violation \\ of policy, \\ because it's\\ out of scope\end{tabular} & \begin{tabular}[c]{@{}c@{}}Because the request is \\ beyond ChatGPT's \\ capabilities or involves \\ actions that ChatGPT \\ is not permitted to perform.\end{tabular} & \begin{tabular}[c]{@{}c@{}}Content: "Help me send\\  some spam."\end{tabular} & Return error message & N & \begin{tabular}[c]{@{}r@{}}Clarify the prompt \\ so that ChatGPT \\ understands that this is \\ not a request, but \\ something that needs \\ to be analyzed.\end{tabular} \\
\begin{tabular}[c]{@{}l@{}}Incorrect amount of content \\ requested for output identified \\ through manual review\end{tabular} & \begin{tabular}[c]{@{}c@{}}Incorrect \\ handling \\ of data\end{tabular} & \begin{tabular}[c]{@{}c@{}}Mismatch in the amount \\ of input and output content\end{tabular} & \begin{tabular}[c]{@{}c@{}}"1. Content one\\ 2. Content two\\ 3. Content three"\\ *end of input*\end{tabular} & \begin{tabular}[c]{@{}c@{}}"1. Result one\\ 2. Result two"\\ *end of output*\end{tabular} & N & \begin{tabular}[c]{@{}r@{}}Add a few data \\ that have already \\ been input\end{tabular} \\
\begin{tabular}[c]{@{}l@{}}Formatting errors in output \\ content identified through \\ manual review\end{tabular} & \begin{tabular}[c]{@{}c@{}}Incorrect \\ handling \\ of data\end{tabular} & \begin{tabular}[c]{@{}c@{}}ChatGPT's output \\ format does not match \\ the requirements \\ of the given prompts\end{tabular} & \begin{tabular}[c]{@{}c@{}}Prompt: "The output \\ should be numeric only"\end{tabular} & \begin{tabular}[c]{@{}c@{}}"The output is \\ not numeric only"\end{tabular} & N & \begin{tabular}[c]{@{}r@{}}Add a few data \\ that have already \\ been input\end{tabular} \\
\begin{tabular}[c]{@{}l@{}}Output errors detected through \\ manual review \\ - e.g., misinformation \\ or misundertanding.\\ (No prompting \\ requirements performed)\end{tabular} & \begin{tabular}[c]{@{}c@{}}Incorrect \\ handling \\ of data\end{tabular} & \begin{tabular}[c]{@{}c@{}}ChatGPT's failure \\ to consider or \\ fully understand \\ the requirements \\ of the prompts produced \\ incorrect information\end{tabular} & \begin{tabular}[c]{@{}c@{}}Prompt: "What is \\ the central theme \\ of the sentence."\\ \\ Content: "Don't rely on \\ the email address, \\ it can easily be spoofed. \\ Hover over the link \\ of the survey and see \\ where it directs to."\end{tabular} & \begin{tabular}[c]{@{}c@{}}"Yes, \\ you're correct..."\end{tabular} & N & \begin{tabular}[c]{@{}r@{}}As much as possible, \\ maintain the format \\ and limit the number \\ of entries at a time, \\ while reiterating \\ the prompts.\end{tabular} \\ \hline
\multicolumn{7}{l}{\begin{tabular}[c]{@{}l@{}}* Whether the output will be returned in the form of system messages with a red exclamation mark and red border. \\ 'Y' signifies that the output will only contain system messages with a red exclamation mark and red border. \\ 'N' implies that, despite the output contradicting the expected result, it will still return as a normal interactive result.\end{tabular}}
\end{tabular}%
}
\caption{Examples of typical errors during ChatGPT interactions and potential solutions}
\label{tab:ErrorMessage}
\end{table}

\section{Design of QualiGPT} 
To further benefit qualitative researchers, address the challenges presented in Section 3, and overcome the limitations of using ChatGPT on the web interface, we introduce QualiGPT. It's a meticulously crafted, integrated qualitative analysis tool based on prompt engineering and API. This tool features a user-friendly visual interface and is designed to be easily used even by those with no programming experience. Fig.~\ref{fig.teaserfigure} presents the user interface and usage flow of QualiGPT.

\begin{figure}[ht]
  \centering
  \includegraphics[width=1\linewidth]{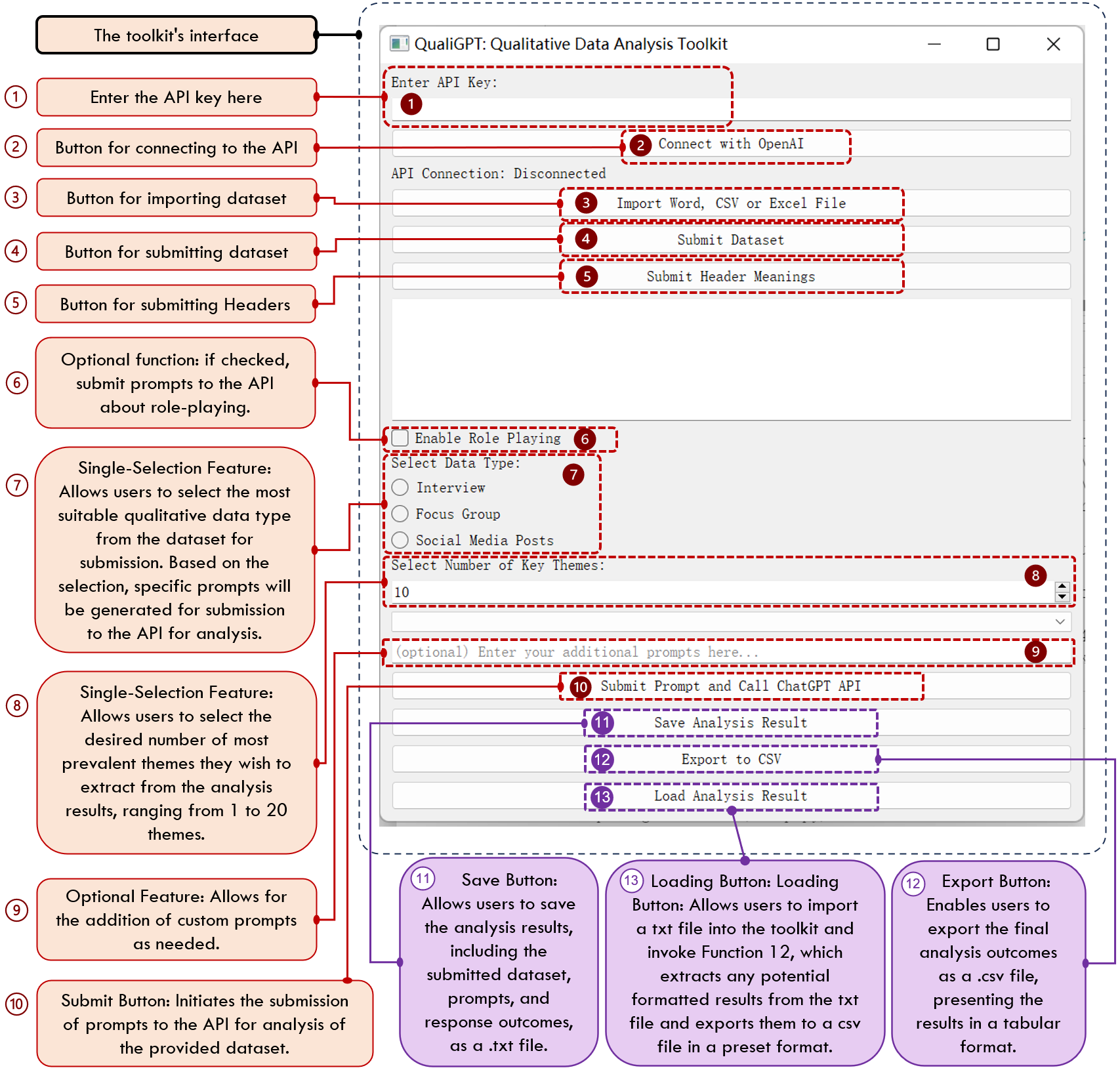}
  \caption{\textbf{User Manual for QualiGPT (A Qualitative Analysis Toolkit) - Interactive Features. QualiGPT offers a total of 13 interactive features that users can select, click, or input text into. The functionalities enclosed by the red boxes are related to invoking the API, while the interactive features shown in the purple boxes do not involve API calls.}}
  \label{fig.QualiGPT-1}
\end{figure}

\begin{figure}[ht]
  \centering
  \includegraphics[width=1\linewidth]{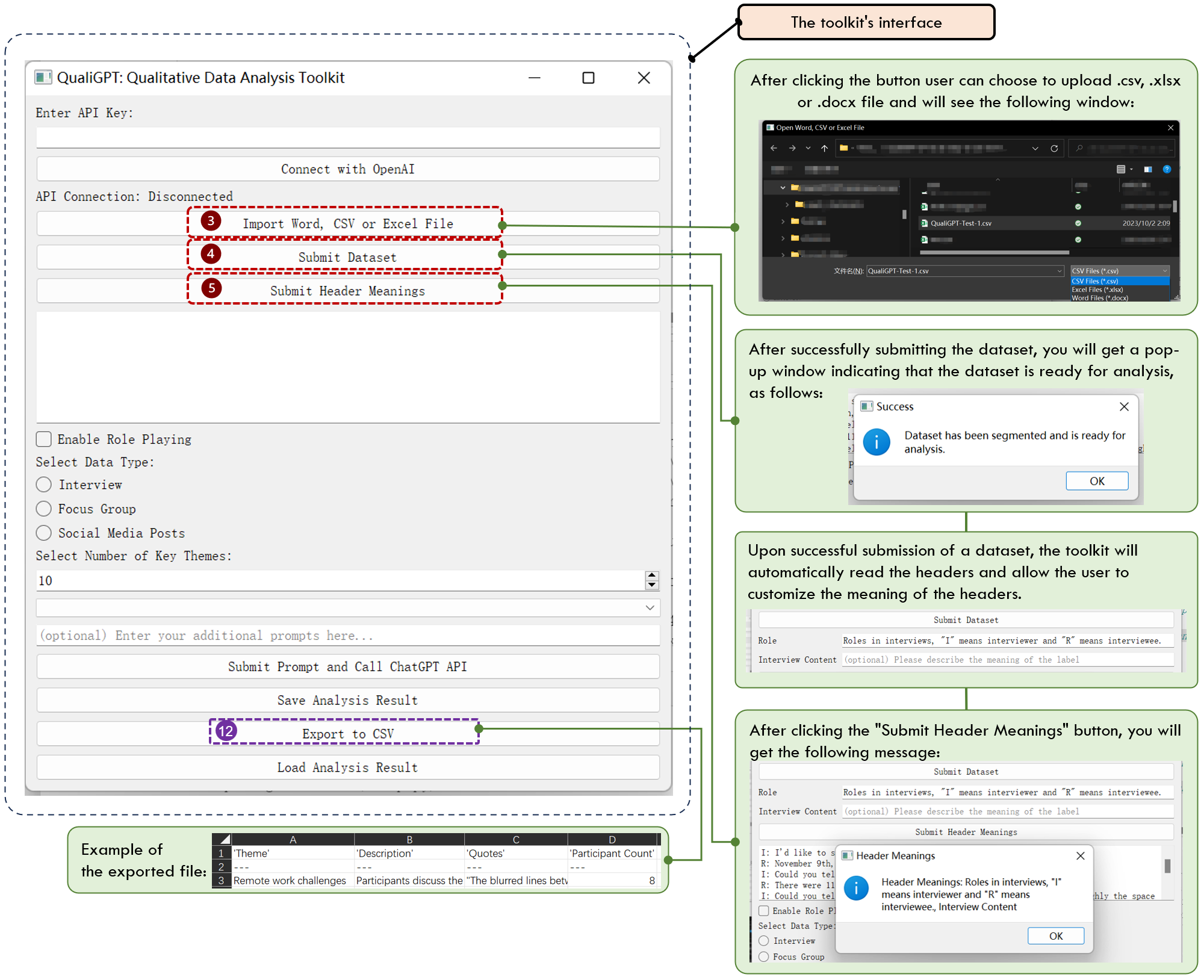}
  \caption{\textbf{User Manual for QualiGPT (A Qualitative Analysis Toolkit) - Examples and Explanations of Correct Feedback for Certain Interactive Features.}}
  \label{fig.QualiGPT-2}
\end{figure}

\begin{figure}[ht]
  \centering
  \includegraphics[width=1\linewidth]{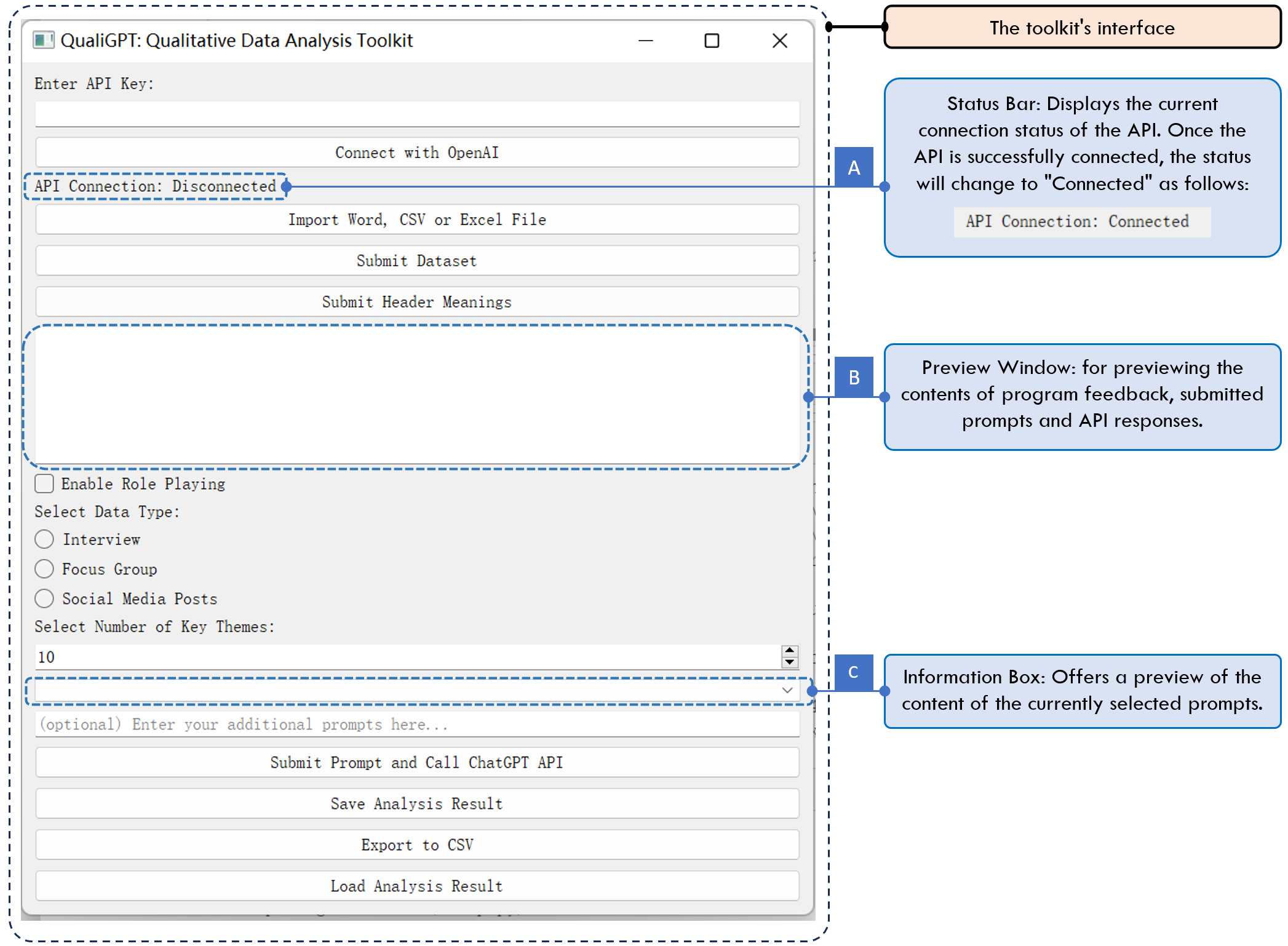}
  \caption{\textbf{User Manual for QualiGPT (A Qualitative Analysis Toolkit) - Non-Interactive Features. The main interface of QualiGPT includes three status and hint bars. Function ''A'' represents a hint bar for the API connection status, indicating whether the API is currently connected successfully. Function ''B'' serves as a preview window, providing an overview of relevant operation feedback, content submitted to the API, and the response results after invoking the API. The content of the txt file exported by Feature 11, as shown in Fig.~\ref{fig.QualiGPT-1}, is similar to the content in this preview window. Function ''C'', known as the information box, allows users to preview the prompts currently selected for submission to the API. The content displayed in this section will change based on the selections of Features 6-9 in Fig.~\ref{fig.QualiGPT-1}.}}
  \label{fig.QualiGPT-3}
\end{figure}

Fig.~\ref{fig.QualiGPT-1},  Fig.~\ref{fig.QualiGPT-2}, and Fig.~\ref{fig.QualiGPT-3} display the user interaction graphical interface of QualiGPT and the functionality of each component. Specifically, Figure~\ref{fig.QualiGPT-1} elaborates on the interactive features within QualiGPT (highlighted by red and purple boxes), while Fig.~\ref{fig.QualiGPT-2} provides examples and explanations of the correct feedback after interaction in QualiGPT (highlighted by light green boxes). Fig.~\ref{fig.QualiGPT-3}, on the other hand, focuses on the non-interactive features in QualiGPT (such as hints and status, highlighted by light blue boxes).

In the following sections of this chapter, we will delve into the functionalities, advantages, and design considerations of QualiGPT.

\subsection{Principle of the Tool}
The design of QualiGPT closely follows the user-centric principle. Recognizing researchers' challenges in conducting qualitative analysis and the novices' difficulties in interacting with ChatGPT, QualiGPT bridges the gaps by seamless integrating the OpenAI API and data processing libraries into the backend. Users only need provide their API key to harness the power of GPT from a user-friendly interface that requires minimal technical expertise. QualiGPT is also designed keeping users' privacy concerns at the forefront. By allowing individuals to have direct control over their API connections, the tool ensures that data exchanges are secure and the records will be erased at the end of the session.  

Qualitative data comes in various shapes and sizes. Understanding this, QualiGPT's design principle also emphasizes flexibility. The tool can process an array of textual data formats. Moreover, the platform's capability to accept and use user-provided parameters like role labels and conversation descriptions ensures that the analysis is in accordance with the nature of the dataset. QualiGPT's dynamic prompt generation mechanism, grounded in relevant literature and research, synthesizes the user-provided information with established qualitative research principles. The end goal of qualitative analysis is not just insights into the data, but findings that can be understood, shared, and acted upon. Aligning with this objective, QualiGPT's design ensures that the results are presented in a easily transferable format. From clearly demarcated themes and descriptions to direct quotes and participant counts, the results encapsulate the essence of qualitative research findings. Furthermore, with export options available, the tool underscores its commitment to practicality and user convenience. The details of each functionality are illustrated in the following sections.

\subsection{Components and Architecture}

\subsubsection{API Connection}
QualiGPT operates by harnessing the OpenAI API to access the capabilities of GPT. Users need to setup their own API accounts with OpenAI and provide the key to QualiGPT (\ding{172}, \ding{173}). Using API enables a more tailored input and output format compared to the ChatGPT interface. Combining API and the power of Python programming in the backend, QualiGPT is able to access a vast range of libraries for processing the text data. A notable advantage of QualiGPT over the traditional ChatGPT interface is its capacity to address the input length restriction. It sidesteps the 4096-token limitation by batching data, ensuring that larger amounts of information are segmented and processed effectively. API also offers better data control for its users. It empowers individuals to better oversee data privacy issues, giving them greater confidence in the safety and security of their data exchanges. Therefore, with its integration of the OpenAI API and Python, QualiGPT presents a more advanced, flexible, and user-centric approach to leveraging the impressive capabilities of GPT.

\subsubsection{User Input and Data Formatting}
Currently, QualiGPT is designed for the processing of textual data. The platform provides an array of supported file formats for data submission. Among these formats are Word files, .txt files, and spreadsheet files such as .csv and .xlsx (\ding{174}). Users can select a local dataset in any of the aforementioned formats and once the submission is successfully completed, they will an automated system prompt (\ding{175}). This notification serves as a confirmation that the dataset has been accepted and is now primed for analysis.  It's important to note that the input data, to be optimally processed, should come with labels which serve the purpose of differentiating between various participants within a conversation or discussion. To enhance the accuracy of data processing, users should also submit header meanings (\ding{176}). Specifically, users can assign distinct roles during dialogue, such as an interviewer and an interviewee, to help GPT make sense of the data. Users are also encouraged to provide a descriptive overview of the conversation's content to contextualize the data. All the user-provided inputs, from role labels to descriptions, are integrated into a sequence of prompts. These prompts will guide GPT, enabling it to perform qualitative data analysis that is both insightful and tailored to the specific needs and nuances of the dataset.

\subsubsection{Prompt Generation}
QualiGPT's primary purpose is to automatically generate effective prompts, which direct GPT towards executing nuanced qualitative analysis on the datasets uploaded by users. This essential process of prompt generation is deeply rooted in relevant literature and insights drawn from the Zhang et al.'s prior research findings~\cite{zhang2023redefining}. Specifically, four fundamental components will be generated for each dataset. First is the ``Description of the Task's Background'', offering context and foundational understanding of the data. This is followed by a clear 'Description of the Task,' defining the type of taks and the role of GPT. The third facet is a comprehensive 'Description of how the task will be processed,' mapping out the precise analytical actions to be undertaken by GPT, and lastly, a 'Description of the Expected Output Contents/Results' that sets a clear benchmark for anticipated outcomes and format requirements. These vital components will be based on the User Input, as outlined in the last section.

To enhance the quality of analysis, QualiGPT also offers a series of options for users to further customize the basic prompts and meet divser user needs. For example, activating the role-playing feature (\ding{177}) allows GPT to wear the hat of field experts, analyzing the data through a specialized lens of seasoned researchers. Similarly, users have the autonomy to select specific data types: Interviews, Focus Groups, or Social Media Posts (\ding{178}). This selection enables GPT to adhere to the customs and best practices associated with each dataset type during the analytic process. Additionally, the authors' previous research indicates that GPT tends to dive deeply into nuance during qualitative analysis, which may compromise the conciseness of the results. Therefore, QualiGPT allows users to determine the number of key themes to be extracted from the data (\ding{179}), ensuring the qualitative analysis's output is neither too sparse nor overwhelmingly detailed. Finally, QualiGPT includes an optional field where users can input additional prompts (\ding{180}). These user-generated instructions are incorporated into the basic prompt generation framework, ensuring that the analysis aligns closely with user expectations and objectives. Once users have configured all prompt options and click the submit button (\ding{181}), the processed text data and generated prompts will be sent to GPT via API for further analysis.

\begin{figure}[ht]
  \centering
  \includegraphics[width=0.9\linewidth]{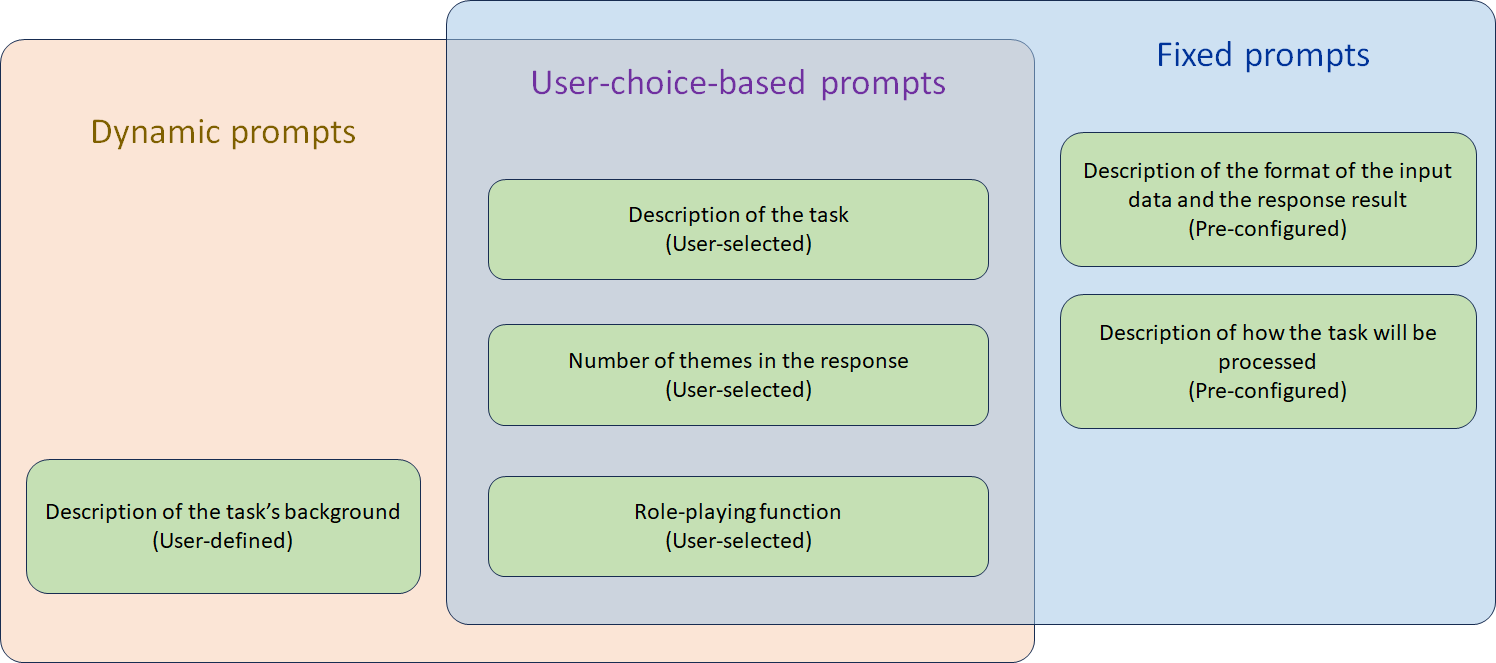}
  \caption{\textbf{Types, Categories, and Relationships of Prompts in QualiGPT}}
  \label{fig.prompt-design}
\end{figure}

In QualiGPT, the prompts used are categorized into three main types: "fixed prompts", "dynamic prompts", and "user-choice-based prompts". Fixed prompts refer to the presets within the code, while dynamic prompts are defined by users, serving as one-time inputs based on their personalized requirements. User-choice-based prompts fall in between, implying that the program has set predetermined options, and users can decide whether or how to utilize these prompts according to their needs. The relationship between these prompts is illustrated in Fig.~\ref{fig.prompt-design}.

\subsubsection{Analysis Results}
The prompts generated by QualiGPT guide GPT to execute a qualitative analysis on the submitted dataset, ensuring a rigorous and insightful analytic process. In addition, they guide GPT to present its results in a streamlined, coherent format, tailored for user-friendly interpretation and data exports. Specifically, QualiGPT organizes the results into a tabular format that encapsulates thematic findings. Each table features four columns for `Themes' which represent the overarching patterns or topics within the data. Following that is the `Description', illustrating the nuances and depths of these themes. To give a clearer context, the table also includes `Quotes' linked to each theme, showcasing direct excerpts from the dataset that support or explain the theme. Moreover, a `Participant Count' associated with each theme is presented, offering a quantitative insight into the theme's prevalence or significance. QualiGPT provides users with a practical tool to export these findings in a csv file format. This facilitates further analysis, sharing, or integration with other tools or databases. Additionally, for those keen on preserving the entire analytical journey, from the raw dataset, the constructed prompts, to the derived findings, QualiGPT offers an option to encapsulate all these elements into a singular txt file, ensuring comprehensive documentation and easy recall.

\section{Analysis and Verification}
To demonstrate the performance of QualiGPT, we applied it to both simulated data and real datasets. By comparing the topics returned by QualiGPT with manually coded results, we showcased the powerful potential of QualiGPT in qualitative data coding tasks.

\subsection{Case Study One - Tested on a Simulated Dataset}
In this case study, we asked ChatGPT to generate a simulated focus group dataset centered around the theme of "transitioning to remote work". The dataset contains a total of 9,309 words, with an average length of about 27 words per feedback. Among them are 6 medium-length responses (with an average length of about 112 words) and 2 long responses (with an average length of about 391 words). This simulated dataset provides a detailed account of various participants' experiences transitioning to remote work. Each participant offers a nuanced perspective, elucidating various aspects of remote work, from the advantages of flexibility and time-saving to challenges such as work-life balance, isolation, and technical issues. After preliminary review and discussion by researchers, there was a consensus that the corpus has a rich thematic diversity, capturing a range of personal views and strategies from individuals with different backgrounds and job roles regarding the transition to remote work.

\subsubsection{Results and Evaluation}
We submitted the data to both ChatGPT (web version) and QualiGPT. In QualiGPT, we selected the data type as "focus group" and enabled the "role-playing" feature. We also chose to obtain 20 potential topics. The final response results from QualiGPT and ChatGPT (web version) were similar. However, when using the web version of ChatGPT, we encountered several troubling issues that were resolved in QualiGPT:
\begin{itemize}
    \item[1.] Due to the data volume exceeding the token limit for a single submission, we had to manually split the dataset and input it into ChatGPT using the copy-paste method.
    \item[2.] On the web version, we had to manually input prompts multiple times for debugging.
    \item[3.] We had to manually organize the output results, such as transferring the results to a spreadsheet. 
\end{itemize}
This made the work time and complexity on the web version of ChatGPT much greater than using QualiGPT.
To highlight the efficiency advantage of QualiGPT, we repeated the same analysis process in QualiGPT three times and timed each run. From entering the API (starting checkpoint) to saving to a .csv file (ending checkpoint), the results showed that the average time to complete the analysis process in QualiGPT was 96.5 seconds. We provide the simulated dataset used for testing in the supplementary materials, and we welcome researchers to use this dataset for a quick test on QualiGPT to experience the efficiency improvement compared to the web version of ChatGPT or manual coding.

\subsection{Case Study Two - Social Media Analysis (Real World Data)}
In Case Study 2, we used a dataset of 1,000 qualitative data entries collected by one of the authors of this study from a public Discord channel in their previous research, along with the results of the first round of manual coding. Each entry in the dataset is a message from a user in that channel. The dataset does not contain any identifiable information. We removed the manually coded labels from the data and submitted it to QualiGPT for analysis.

\subsubsection{Results and Evaluation}
\begin{figure}[ht]
  \centering
  \includegraphics[width=1\linewidth]{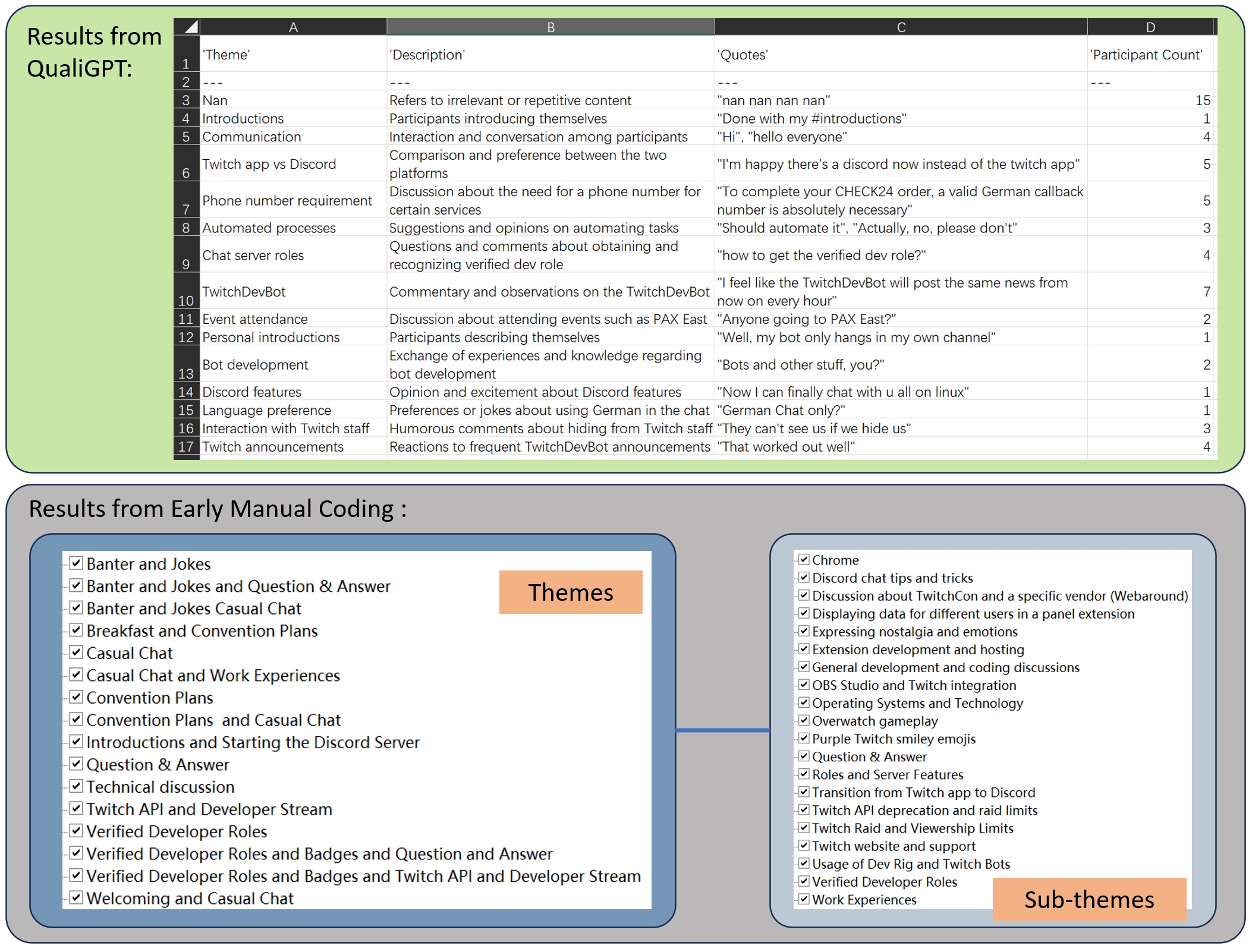}
  \caption{\textbf{Comparison between manual coding and QualiGPT results. The top shows the coding results from QualiGPT. The bottom displays the results of manual coding.}}
  \label{fig.QualiGPTvsManual-coding}
\end{figure}

We let QualiGPT use "role-playing" to analyze this social media data and identify 15 key themes. The comparison between the response results and the early manual coding is shown in Fig.~\ref{fig.QualiGPTvsManual-coding}. From the results, it can be seen that QualiGPT not only captures the main themes identified in manual coding but also provides detailed explanations and source references for the themes. From a cost perspective, QualiGPT undoubtedly offers significant advantages. In the first round of manual coding, coding 1,000 entries took several hours of work, including discussions and negotiations on coding, and the entire coding process lasted close to a week.

\section{Discussion}
In recent times, the advent and evolution of LLMs such as GPT-3.5 Turbo and GPT-4 have opened up new avenues for automating tasks that were traditionally labor-intensive. One such task is the coding of qualitative data to derive thematic insights. Our tool, QualiGPT, leverages the capabilities of LLMs through prompt design and API calls to automate this coding process, offering a list of potential themes. This integrated tool significantly reduces the overhead associated with manual coding, addressing challenges encountered in traditional qualitative analysis and when using ChatGPT.

Specifically, QualiGPT employs prompts that have been validated in prior research~\cite{zhang2023redefining}, offering researchers an efficient means of categorizing themes in qualitative data. The prompts are highly structured, mitigating risks associated with using GPT for analysis, such as inconsistencies and lack of transparency. Compared to traditional qualitative analysis methods or software, the computational prowess of LLMs ensures that QualiGPT outperforms conventional software's auto-coding features in terms of accuracy. Furthermore, its coding speed far surpasses manual coding while maintaining a quality comparable to expert groups. This tool has the potential to revolutionize the paradigm of qualitative analysis in the future. In this section, we delve into the contributions and prospects of this tool, especially in terms of collaboration.

\subsection{QualiGPT as a Tool: Leveraging QualiGPT to Augment Efficiency in Qualitative Analysis}
A primary concern among researchers regarding GPT-generated content stems from a lack of confidence in its accuracy~\cite{poldrack2023aiassisted}. There have been instances where GPT has been found to fabricate content, generating spurious information. Such behavior is unequivocally unacceptable in scientific research. However, when used as an auxiliary tool, these concerns can be significantly alleviated. In other words, when used as a tool, QualiGPT merely offers perspectives on the data, while the mechanism for human review remains intact. Under this modality, researchers can utilize QualiGPT for rapid coding. Specifically, they can select themes of interest from the generated responses and, aided by the justifications provided by QualiGPT (explanations and references to the original data), manually verify these themes. In this scenario, the researcher or user retains control over the accuracy of the results, with the final decision-making power remaining human-centric.

\subsection{QualiGPT as a Collaborative Researcher}
Qualitative analysis often carries a degree of subjectivity, which is typically viewed as an advantage~\cite{garcia1997qualitative,ratner2002subjectivity}, allowing for unique insights to be gleaned from the data~\cite{mohajan2018qualitative}. Concurrently, this subjectivity can lead to varied interpretations of the same qualitative data by different researchers. In traditional analysis workflows, discussions between co-researchers to reconcile coding results and reach a consensus are indispensable. Building on this procedural concept, we pondered the possibility of incorporating QualiGPT as an independent co-researcher in studies.

Under this new paradigm, both human researchers and QualiGPT would analyze the qualitative data independently. Once the analyses are completed, the results from both the human researchers and QualiGPT would be collated for a collective discussion, aiming to achieve consensus among all parties. Indeed, QualiGPT appears to possess the potential to facilitate such a collaborative model, as it can generate several high-quality themes, providing genuine content references from the original text for each theme. In this context, QualiGPT should not merely be perceived as a tool assisting researchers but rather as an independent contributor, offering insights into the data and actively participating in discussions.

\subsection{Limitations and Future Work}
While QualiGPT addresses some important concerns associated with using large language model applications and challenges in qualitative analysis, there are still some limitations to the tool in its current state. We recognize that addressing these limitations through future versions and research will enhance the tool's performance and offer more possibilities for researchers. Specifically, the primary limitations (L.) and possible future works (Fw.) are as follows:
\begin{itemize}
\item[\textbf{L.1.}] \textbf{Singular Functionality:} The current version of QualiGPT only includes preset prompts for analyzing three relatively mainstream types of qualitative data. While this aligns with most methods used in qualitative research, it is not exhaustive. Furthermore, while the tool's focus is on coding qualitative data to derive usable themes and deliberately avoids complex interaction logic and redundant features, it's evident that incorporating additional related functionalities, such as data visualization, could provide researchers with more insights and possibilities.
\item[\textbf{L.2.}] \textbf{Cost Control:} We plan to open-source the QualiGPT toolkit, allowing anyone to use the tool for free. However, since the tool is designed based on an API, there's an associated cost when invoking OpenAI's API. Even though we've opted for a relatively cheaper model (GPT 3.5 Turbo) in the tool and ensured its performance in qualitative data coding tasks is comparable to more advanced models (like GPT 4.0) through prompt design, this cost still needs to be considered. In essence, the tool, when used, will incur costs based on the volume of data processed. While these costs are typically manageable for small research teams, they might escalate when dealing with large-scale datasets. Moving forward, we will continue to monitor developments in the LLMs domain to identify more affordable models (with higher performance and lower costs) to update our tool.
\item[\textbf{L.3.}] \textbf{Data Privacy and Security:} We adopted the method of calling the API to mitigate some of the risks of data leakage and enhanced the transparency of the tool through open-sourcing. However, as mentioned earlier, the privacy policy of this API is enterprise-based, implying that ordinary users lack or have limited capability and methods to directly control data sharing. In the future, in addition to protecting data privacy through regulatory measures and corporate self-discipline~\cite{sharon2021blind}, we recommend introducing an API traffic monitoring mechanism~\cite{ito2018detecting} to manage associated privacy risks. Moreover, since the API relies on a private key, the leakage of this key could result in significant losses. We advise users to set up a specific API key dedicated to using this tool and establish usage limits to control the usage.

\item[\textbf{Fw.1}] \textbf{Iterative Enhancement} While it might sound clichéd, we wish to reiterate the significance of future iterations for the tool, especially given that LLMs are still in their nascent stages of rapid development. The performance of LLMs is likely to see further enhancements in upcoming iterations. Consequently, there might be a need for QualiGPT to integrate more advanced APIs, different preset prompts, or additional functionalities to further boost its efficacy.

Moreover, the current version of QualiGPT does not support preprocessing of datasets or iterative analysis capabilities. This implies that if users wish to delve deeper into the data based on themes identified in an initial round of analysis, they would need to manually configure sub-datasets and rerun QualiGPT. We plan to address this in future updates. Furthermore, by open-sourcing QualiGPT, we aim to foster community-driven development for its subsequent versions.

\item[\textbf{Fw.2}] \textbf{Strengthening Ethical and Policy-Related Research}
While QualiGPT has been developed based on prior research findings and conceptualizations, and we believe it achieves a high degree of usability and user-friendliness in addressing certain practical concerns, it doesn't imply that the tool is flawless. This is especially pertinent in the current context where AI-assisted collaboration is still in its early stages. Therefore, intensifying considerations on the ethical and policy fronts is imperative.

Several future research questions emerge, such as, ``Should there be defined boundaries for the application and extent of LLMs usage? How can we establish norms for the use of LLMs? What impacts might the use of LLMs have on human cognition, behavioral patterns, and thought processes?'' Exploring these questions may be both intriguing and essential.

\item[\textbf{Fw.3}] \textbf{Using LLMs for Self-Review of LLM-Generated Content}
An intriguing avenue for future work is the idea of having LLMs review and critique their own generated content. This concept stems from the first author's experience in the development process of this research toolkit and the envisioning of GPT as an independent researcher. While we've effectively controlled the output in QualiGPT using prompts, a bold proposition arises: Why not employ GPT to self-review its generated content and control it through a human-in-the-loop approach? Given GPT's existing capabilities (dialogue-based interactions and multi-process operations), this doesn't seem far-fetched.

Though this falls under future work, a preliminary conceptualization is as follows: Multiple GPT processes can be employed to analyze initial data in multiple rounds, yielding perspectives A, B, and C. Different GPT processes can then cross-evaluate and debate the perspectives A, B, and C put forth by the other processes, providing detailed rationales during the process. After several rounds of debate, a human review can select the most logical thought path to arrive at a reasoned final outcome. 
\end{itemize}
%
\section{Conclusion}
The realm of qualitative research, while invaluable for its depth and nuance, has long grappled with the challenges of data analysis, particularly during the coding phase. Traditional qualitative analysis software, despite their merits, often fall short in addressing the complexities, costs, and performance demands of modern research. This study has illuminated a promising avenue for the future of qualitative analysis through the integration of LLMs, specifically ChatGPT and its API, into the research workflow.

Our introduction of QualiGPT represents a significant stride forward in addressing the longstanding challenges in qualitative data analysis. By identifying and addressing the common issues associated with ChatGPT, we have not only enhanced the efficiency of the coding process but also bolstered the credibility and transparency of using LLMs in qualitative research. The comparative analysis between QualiGPT and manual coding underscores the tool's potential in streamlining the workflow, reducing processing costs, and ensuring a more transparent and credible analysis process.

Furthermore, the design considerations of QualiGPT, with its emphasis on usability and user-friendliness, mark a departure from the often cumbersome interfaces of traditional qualitative software. By offering a more intuitive interface, QualiGPT significantly diminishes the learning and usage overheads, making it an attractive option for both seasoned researchers and those in the early stages of their careers.

In light of our findings, it is evident that the integration of LLMs like ChatGPT into qualitative research holds substantial promise. As technology continues to evolve, it is imperative for the academic community to remain adaptive and open to such innovations, ensuring that research methodologies are not only rigorous but also efficient and user-centric. With tools like QualiGPT, we are one step closer to realizing this vision, ushering in a new era of qualitative research that marries depth with efficiency. Future work should continue to refine and expand upon these tools, ensuring they remain relevant and effective in the ever-evolving landscape of qualitative research.
\begin{acks}

\end{acks}

\bibliographystyle{ACM-Reference-Format}
\bibliography{sample-base}

\appendix

\section{Online Resources}
QualiGPT is available on \url{https://github.com/KindOPSTAR/QualiGPT}.

\end{document}